\documentclass[prl,showpacs,floatfix,twocolumn,amsmath,amssymb,dvips]{revtex4-1}

\usepackage{epsfig}
\usepackage{graphicx}
\usepackage{dcolumn}
\usepackage{bm}
\usepackage[colorlinks=true,dvipdfm]{hyperref}

\begin{document}


\title{Complete scaling makes differences along the critical isobar:
Molecular dynamics simulations of Lennard-Jones fluid with finite-time scaling}

\author{Yankui Wang}
\author{Ji Liu}
\author{Jiajie Xiao}
\author{Fan Zhong}
\thanks{Corresponding author. Email: stszf@mail.sysu.edu.cn}
\affiliation{State Key Laboratory of Optoelectronic Materials and Technologies, School of Physics and Engineering, Sun Yat-sen University, Guangzhou 510275, People's Republic of China}

\date{\today}

\begin{abstract}
We show that, along the critical isobar, the complete scaling results in a unique leading scaling qualitatively distinct to that arising from the simple and the revised scalings. This is verified by a complete-field finite-time scaling theory, which combines the complete scaling with finite-time scaling, and its application to the molecular dynamics simulations of the vapor-liquid critical point of a three-dimensional one-component Lennard-Jones fluid in an isobaric-isothermal ensemble with linear heating or cooling. Both the static and the dynamic critical exponents as well as the critical parameters can be estimated without \emph{a priori} knowledge of the universality class. The results agree with extant values and thus show the necessity of the complete scaling to the leading asymptotic behavior along the critical isobar even for the LJ fluid whose asymmetry is thought to be weak.
\end{abstract}

\maketitle

Fluid criticality still attracts great attention though it was first recorded nearly one and a half century ago~\cite{Andrews}. A primary concern is the lack of symmetry between the vapor and the liquid phases for usual fluids and the relevant complicated field mixing~\cite{bsa}.

In the ``simple scaling'', no mixing is needed and the reduced temperature $\tau=(T-T_{c})/T_{c}$ and the chemical potential $\hat{\mu}=(\mu-\mu_{c})/k_{\rm B}T_{c}$ correspond directly to the thermal and the ordering fields in the Ising model, where $k_{\rm B}$ is the Boltzmann constant and the subscript $c$ stands for the values at the critical point hereafter. This leads to the usual singularity of the order parameter $\Delta\hat{\rho}=(\rho-\rho_{c})/\rho_{c}\sim|\tau|^{\beta}$ with its critical exponent $\beta$, where $\rho$ is the density. Yet, along the vapor-liquid coexistence curve this leading behavior cancels and the resultant diameter $\rho_d=(\rho_{\rm vap}+\rho_{\rm liq})/2=\rho_c$ is symmetric, different from the asymmetry in the empirical law of rectilinear diameter $\rho_d\propto|\tau|$~\cite{lrld}. In the ``revised scaling''~\cite{rm}, mixing of the two physical fields gives rise to a subleading term in $\Delta\hat{\rho}$, a term which accounts for the $|\tau|^{1-\alpha}$ singularity in $\rho_d$~\cite{Widom70,Mermin}, where $\alpha$ is the critical exponent of the isochoric heat capacity.
Recently, a ``complete scaling'' has been proposed~\cite{fish00,kfo03,ofp,kf} in which the pressure $P$ is further mixed into the scaling fields in order to account for the Yang-Yang (YY) anomaly~\cite{yy}. This results in yet another term in $\Delta\hat{\rho}$ and hence an additional $|\tau|^{2\beta}$ singularity in $\rho_d$, which is dominant as $2\beta<1-\alpha$ normally. The complete scaling is supported by data on vapor-liquid coexistence and heat capacity in highly asymmetric fluids~\cite{aniwang}, and is found to be equivalent to a field-theoretic treatment of asymmetric fluid criticality up to independent fifth-order terms~\cite{Bertrand}. Some experiments from binary fluids also support the $2\beta$ singularity~\cite{Cerd,Wang08,Perez15,MHuang,Vale} and the YY anomaly~\cite{Perez21,Losada}. However, experimental evidences from specific heat~\cite{ofu} for the YY anomaly may stem from small traces of impurities~\cite{Wycz}, and the two singular terms may compensate in some fluids, resulting in an almost rectilinear diameter~\cite{aniwang}. As the leading singularity of $\Delta\hat{\rho}$ remains intact, a question then arises as to whether it could be changed so that the complete scaling could be more detectable.

Another concern is computer simulations. Although simulational estimates of the critical properties of lattice systems such as Ising models and the like has reached a high level of accuracy, that of off-lattice systems such as fluids is less satisfactory. Besides the asymmetry, another main difficulty arises from the divergent correlation length $\xi$ for fluctuations at critical points. Usual methods such as direct interfacial~\cite{Widom,Gubbins}, Gibbs ensemble Monte Carlo (MC)~\cite{panagio}, isobaric-isothermal ensembles ($NPT$, $N$ is particle number)~\cite{mf,okumura} or canonical ($NVT$, $V$ is volume)~\cite{Szalai,okumura01} ensembles plus test particle~\cite{widom} usually work away from the critical point due to strong fluctuations there and the accompanying extrapolation to determine critical properties is questionable~\cite{wilding,wahu}.

Near to critical points, finite-size scaling (FSS)~\cite{fss,Cardyfss,privman} is crucial due to the divergent $\xi$. MC simulations in grand canonical ($\mu VT$) ensembles, amenable to an FSS analysis, are thus usually employed~\cite{wilding}. By contrast, $NPT$ ensembles need a revised FSS and are deemed to be computationally less efficient~\cite{Wildingb}. A mixed-field FSS theory~\cite{Bruce,Wilding92,wilding} based on the revised scaling for estimating critical properites of fluids has been developed using $\mu VT$ ensemble MC simulations with histogram reweighting~\cite{ferswe}. Yet, this method is ``biased'' in that \textit{a priori} knowledge of the Ising universality class ought to be given~\cite{kim}. An unbiased recursive algorithm based on the complete scaling using the FSS of cumulants~\cite{Binderc} has also been developed again with MC simulations in $\mu VT$ ensembles~\cite{kfl,kflc,Kim05}. However, the method demands a significant amount of high quality precise data and adjusting parameters to be determined to exact or approximated scaling functions. In addition, dynamic critical behavior has yet to be incorporated.

When dynamics is considered one encounters the divergent relaxation time $t_{\rm eq}$ that brings about critical slowing down as critical points are approached. Also, critical dynamics of fluids is complicated by hydrodynamic modes and is described by model H~\cite{hohenberg}. Consequently, although there exist consistent theoretical~\cite{ohka,sighh,hfb} and experimental~\cite{burssengers,bergzimmerli} results, molecular dynamics (MD) simulations for the critical dynamics of a Lennard-Jones (LJ) fluid~\cite{chen} and a Widom-Rowlinson mixture~\cite{jy} appeared merely about a decade ago. The former used a specially designed, near equilibrium initial condition in an $NVT$ ensemble but knowledge of equilibrium critical properties is a prerequisite. The latter employed an event-driven MD algorithm to expedite the dynamics~\cite{Smith}. However, the value of the dynamic critical exponent $z$ obtained disagrees with the accepted ones~\cite{sm}. A correction by background remedies the discrepancy but theoretical values have to be invoked using microcanonical ensemble MD simulations~\cite{das}. In the latter two pieces of work, semi-$\mu VT$ ensembles~\cite{deM} are employed to provide both equilibrium properties and initial conditions for the subsequent MD calculations. One sees that MD simulations alone could not yet produce all critical properties though they simulate the real dynamic process.

Here, we show that the complete scaling can result in a qualitatively distinct \emph{leading} behavior to the other two scalings along the critical isobar for $\Delta\hat{\rho}$ itself and other thermodynamic functions besides $\rho_d$. As $P$ is mixed into the scaling fields in the complete scaling, if we fix it and approach the critical point along the critical isobar, both scaling fields are affected, in sharp contrast to the other two scalings. To show this, we explore the \emph{time} aspect and develop a theory of ``complete-field'' finite-time scaling by combining finite-time scaling (FTS)~\cite{gong,zhong} with the complete scaling. Moreover, we are able to effect the theory with MD simulations of a three-dimensional pure LJ fluid. $NPT$ ensembles are employed though $T$ is varied linearly with time to realize FTS, a situation which can be realized in experiments. Our numerical results support to the leading asymptotic behavior the complete scaling, which is thus needed along the unique critical isobar even for the LJ fluid whose asymmetry is believed to be weak. We can also determine both the static and dynamic critical exponents as well as the critical parameters with the dynamic $NPT$ ensembles~\emph{alone}.

FTS has been applied effectively both to classical~\cite{gong,huang,xiong,xiong12,Huang,zhong} and quantum phase transitions in lattice models~\cite{Yin13,Yin14,Hu}. Analogue to FSS that circumvents the problem that $\xi$ is longer than the lattice size $L$, FTS overcomes the critical slowing down of a divergent $t_{\rm eq}$ by devising a controllable time scale $t_{\rm dr}$. This $t_{\rm dr}$ is obtained by linearly varying $T$ at a constant sweep rate $R$ to drive a system through its critical point. Although $t_{\rm dr}$ is shorter than $t_{\rm eq}$ and thus the system falls out of equilibrium in the FTS regime, the FTS enables one to extract both static and dynamic critical properties similar to its spatial counterpart. Moreover, in the FTS regime, the effective length scale arising from the driving is shorter than $L$ and thus finite-size effects are only subsidiary and negligible.


To see this, note that the time evolution of the characteristic thermodynamic function in an $NPT$ ensemble under a coarse-graining of a factor $b$ is
\begin{equation}
\tilde{\mu}=b^{-\beta\delta/\nu}\tilde{\mu}(\tilde{\tau} b^{1/\nu},\tilde{p}b^{(2-\alpha)/\nu},tb^{-z}),\label{mpttbs}
\end{equation}
where $\tilde{\mu}$, $\tilde{\tau}$, and $\tilde{p}$ are the scaling fields corresponding to $\hat{\mu}$, $\tau$, and $\hat{p}=(P-P_{c})/\rho_{c}k_{\rm B}T_c$, respectively, and $\delta$ and $\nu$ are the critical exponents of the critical isotherm and $\xi$, respectively. For simplicity of presentation, we shall use identical symbols for all the three scalings and differentiate the latter by their contents. Thus, in the simple scaling, all the three scaling fields are just the physical fields themselves, while in the complete scaling in the linear approximation~\cite{fish00,kfo03,aniwang},
\begin{equation}
\tilde{p}=\hat{p}-k_{0}\tau-l_0\hat{\mu}, \quad \tilde{\tau}=\tau-l_{1}\hat{\mu}-j_{1}\hat{p},\quad\tilde{\mu}=\hat{\mu}-k_{1}\tau-j_{2}\hat{p},
\label{cssf}
\end{equation}
where all coefficients are constants. In the revised scaling, $k_0$, $l_0$, $j_1$, and $j_2$ all equal zero.
In order to avoid reaching an apparently different universality class, we have associated the field variable $\tilde{p}$ and the function $\tilde{\mu}$ in the $NPT$ ensemble with the exponents $2-\alpha$ and $\beta\delta$, respectively, a reversal of the usual assignment in magnetic systems~\cite{Cardy}.

In the simple scaling, Eq.~(\ref{mpttbs}) describes the usual critical dynamics; but it is valid even if $\tau=Rt$, viz., time dependent, which has been established by a dynamic renormalization-group (RG) theory~\cite{zhong06,gong,zhong}. Replacing $\tau$ with $R$ and choosing $b$ such that $Rb^{r_T}$ is a constant, one arrives at an FTS form
\begin{equation}
\hat{\mu}=R^{\beta\delta/r_T\nu}f_1(\hat{p}R^{-(2-\alpha)/r_T\nu},tR^{z/r_T}),\label{mptsfts}
\end{equation}
where $f_i$ ($i$ an integer) is a scaling function and $r_T$ is the RG eigenvalue of $R$. We have neglected  dimensional factors and the possible difference in $f_i$ between the two phases hereafter for simplicity. $r_T=z+1/\nu$ from $\tau=Rt$ and its coarse-grained version~\cite{zhong06}. Equation~(\ref{mptsfts}) shows that the driving imposes a timescale $t_{\rm dr}\sim R^{-z/r_T}$ on the evolution of the system. In the FTS regime, $tR^{z/r_T}\ll1$, or, back to $\tau$, $t_{\rm eq}\sim|\tau|^{-\nu z}\gg t_{\rm dr}$ as expected. If one includes $L$ in Eq.~(\ref{mpttbs}), one will see that the FTS regime is given additionally by $L\gg R^{-1/r_T}$, which is an effective driven length scale. So, a series of driving circumvents both the divergent length and the divergent time.

Next we turn to the critical isobar and develop the complete-field FTS theory by incorporating the complete scaling in FTS. Note that all scaling fields vary with $\tau$ and the method that leads to Eq.~(\ref{mptsfts}) does not work. So, we need to find the leading behaviors of all the scaling fields with $\tau$ at $P=P_c$. This can be done by expanding the equilibrium scaling form,
\begin{equation}
\tilde{\mu}=|\tilde{\tau}|^{\beta\delta}f_2(\tilde{p}/|\tilde{\tau}|^{2-\alpha}), \label{htao}
\end{equation}
which results from Eq.~(\ref{mpttbs}), in $\tilde{p}$ for vanishingly small $\tilde{\tau}$ and using (\ref{cssf}), resulting to the leading order in~\cite{kfo03}
\begin{equation}
\tilde{\tau}\sim\tau,\qquad\tilde{\mu}\sim\tilde{\tau},\qquad\tilde{p}\sim|\tilde{\tau}|^{(2-\alpha)/\beta\delta}.
\label{phtau}
\end{equation}
Since $\Delta\hat{\rho}=(\partial\hat{p}/\partial\hat{\mu})_{\tau}-1$, it can be related to the scaled density $(\partial\tilde{p}/\partial\tilde{\mu})_{\tilde{\tau}}$ through Eq.~(\ref{cssf}). So,
\begin{equation}
\Delta\hat{\rho}=b^{-\beta/\nu}\Delta\hat{\rho}(\tilde{\tau} b^{1/\nu},\tilde{p}b^{(2-\alpha)/\nu},tb^{-z}),\label{rhotpt}
\end{equation}
similar to Eq.~(\ref{mpttbs}) to the leading asymptotic term~\cite{fish00,kfo03,aniwang}. Substituting Eq.~(\ref{phtau}) into (\ref{rhotpt}), choosing a scale such that $tb^{-z}$ is a constant, and expressing $t$ in terms of $\tau$ and $R$, we finally obtain the leading singularity
\begin{eqnarray}
\Delta\hat{\rho}(\tau,R)&=&
(\tau/R)^{-\beta/\nu z}f_{3}(\tau R^{-1/r_T\nu},\tau R^{-\beta\delta/r\nu})\nonumber\\
&\simeq&R^{\beta/r\nu}f_{4}(\tau R^{-\beta\delta/r\nu}),\label{drho}
\end{eqnarray}
where $r=z+\beta\delta/\nu$ and in comparison to the neglected subleading terms we have neglected in the second line a correction to scaling since ${\beta\delta/r\nu-1/r_T\nu}>0$. Note that $r$ is the rate exponent associated with varying the ordering field in magnetic systems~\cite{zhong06,gong,zhong}. This can also be seen if we consider instead a $\mu VT$ ensemble and vary $\hat{\mu}$ in the simple scaling (see below).

The FTS form~(\ref{drho}) is qualitatively distinct from those for the simple and the revised scalings in the same process. In the simple scaling, Eq.~(\ref{mptsfts}) results obviously in an FTS along the critical isobar characterized by the \emph{thermal} rate exponent $r_T$ instead of the \emph{field} rate exponent $r$. In the revised scaling, $\tilde{p}=\hat{p}=0$ then leads to $\tilde{\mu}\sim|\tilde{\tau}|^{\beta\delta}$ from Eq.~(\ref{htao}). This then gives rise to $\hat{\mu}\sim\tau$ and so $\tilde{\tau}\sim\tau$ from Eq.~(\ref{cssf}). So, from Eq.~(\ref{rhotpt}), varying $\tau$ can only result in an FTS form similar to the simple scaling.

We now show that the critical isobar with its result is unique in the sense that along other paths such as the coexistence curve or the critical isokyme, the leading behavior of the complete scaling cannot be uniquely distinguished from the others. Along the coexistence curve, chosen as $\tilde{\mu}=0$~\cite{aniwang}, $\tilde{p}\sim|\tilde{\tau}|^{2-\alpha}$ from Eq.~(\ref{htao}). As a result, in Eq.~(\ref{rhotpt}), $\tilde{p}b^{(2-\alpha)/\nu}$ is a function of $\tilde{\tau} b^{1/\nu}$ and thus only $r_T$ matters. So do the other two scalings. Along the critical isokyme, viz. fixing $\mu$ to $\mu_c$ and varying $T$, Eq.~(\ref{phtau}) is also valid~\cite{kfo03} and thus Eq.~(\ref{drho}) applies as well. This is also true for the revised scaling, though in the simple scaling, only $r_T$ appears as both paths coincide.

Note that the above conclusions do not depend on the ensemble used. In a $\mu VT$ ensemble, $\tilde{p}b^{(2-\alpha)/\nu}$ in Eq.~(\ref{rhotpt}) is replaced by $\tilde{\mu}b^{\beta\delta/\nu}$. Along the critical isobar, for the complete scaling, the first two expressions in~(\ref{phtau}), valid in this case too, leads again to~(\ref{drho}). For the revised scaling, again $\tilde{\mu}\sim|\tilde{\tau}|^{\beta\delta}$ and $\tilde{\tau}\sim\tau$. Accordingly, both $\tilde{\mu}b^{\beta\delta/\nu}$ and $\tilde{\tau}b^{1/\nu}$ are reduced to $\tau b^{1/\nu}$ and thus the FTS form involves $r_T$ instead of $r$, similar to the simple scaling. Note also that the results are not due to FTS. In equilibrium, one can see that the complete scaling leads to a leading behavior of $\Delta\hat{\rho}\sim|\tau|^{1/\delta}$~\cite{kfo03}, whereas the other two result in $|\tau|^{\beta}$, reflecting both the mixing of $P$ and the dominating of the field.

Having known the FTS form of $\Delta\hat{\rho}$, those for other observables can be similarly derived. For example, the isobaric specific heat $C_p$ and the reduced entropy density $\Delta\hat{S}=(S-S_{c})/\rho_{c}k_{B}$ along the critical isobar, the former calculated from the fluctuations of enthalpy and the latter from the integration of the former, behave as
\begin{eqnarray}
C_{p}(\tau,R)-C_{p0}&\simeq& R^{-\gamma/r\nu}f_{5}(\tau R^{-\beta\delta/r\nu}),\label{cpsca1}\\
\Delta\hat{S}(\tau,R)&\simeq& R^{(1-\alpha)/r\nu}f_{6}(\tau R^{-\beta\delta/r\nu}),\label{ssca1}
\end{eqnarray}
respectively, to the leading singularity, where $C_{p0}$ is a background contribution.

From the scaling forms, one can then determine the critical properties using the FTS methods~\cite{gong,zhong}. We utilize $\rho$ and $S$ instead of the peaks in $C_p$ as they are smoother. If $P_c$ has been known, one can use Eqs.~(\ref{drho}) and (\ref{ssca1}) and a method of iterations to estimate the parameters involved. In particular, given an initial $T_{c}$, one obtains $\rho$ at different sweep rates at this $T_{c}$. Then $\rho_{c}$ and $\beta/r\nu$ can be extracted
from
\begin{equation}
\rho=\rho_{c}+a_{1}R^{\beta/r\nu},\label{opextra}
\end{equation}
according to Eq.~(\ref{drho}), where $a_i$ ($i$ an integer) is a constant. At this $\rho_{c}$, the temperature $T$ at each rate can also be identified to find a new $T_{c}$ and $\beta\delta/r\nu$ from
\begin{equation}
T=T_{c}+a_{2}R^{\beta\delta/r\nu}\label{extractedtc}
\end{equation}
again from Eq.~(\ref{drho}). Several iterations suffice to reach convergent $T_{c}$ and $\rho_{c}$ and thus $\beta\delta/r\nu$ and $\beta/r\nu$. A similar procedure can be conducted for $S$ and thus $S_c$, $(1-\alpha)/r\nu$, and again $T_c$ and $\beta\delta/r\nu$ can be obtained, the latter two parameters can serve as a consistency check.

We have yet to determine $P_c$. Note that at a critical point, $T_c$ obtained from the peak temperatures of $C_{p}$ using Eq.~(\ref{extractedtc}) during heating and cooling is identical; however, it is different below $T_c$~\cite{Chu,zhong05}. This can be employed to estimate $P_c$~\cite{xiong}.

Our model system contains atoms interacting via the LJ potential $\phi(x)=4\epsilon[(\sigma/x)^{12}-(\sigma/x)^6]$,
where $x$ is the distance between two atoms and $\epsilon$ and $\sigma$ specify the units of energy and length, respectively. Thus, we use reduced quantities denoted by asterisks such as energy $E^{*}=E/\epsilon$, time $t^{*}=t/(m\sigma^2/\epsilon)^{1/2}$, $L^{*}=L/\sigma$, $T^{*}=k_BT/\epsilon$, and $P^{*}=P\sigma^{3}/\epsilon$, where $m$ is the mass of an atom. The full LJ potential is used with a cutoff  at $3.0\sigma$. Berendsen's method is applied to couple $T^*$ and $P^*$ of the system to the bath~\cite{berendsen}. We employ the leap-frog algorithm~\cite{frensmit} with a time step $\delta t^*=0.001$. The system contains $150$ atoms. At each given $P^*$, the system is heated or cooled from an equilibrium state to the other state by linearly varying $T^*$ at a series of prescribed $R^*$ from $5\times10^{-10}$ to $1\times10^{-8}$. Larger rates result in undistinguishable evolution but lower rates are time consuming and may cross over to an FSS regime~\cite{zhong,Huang}. Several thousand samples with different initial states are run for average. At $P^*=0.140$, only heating is conducted and the sample size is $10,000$.

\begin{figure}
\includegraphics[height=3.5cm]{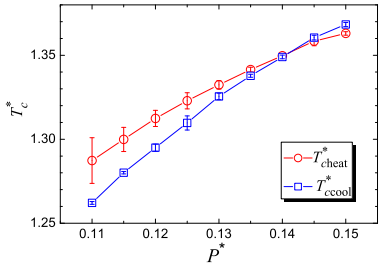}
\caption{\label{pc}(Color online) $T_{c}^{*}$ extracted from Eq.~(\ref{extractedtc}) at various $P^*$ from heating and cooling. Lines are a guide to the eye.}
\end{figure}
Figure~\ref{pc} shows the estimation of $P_c^*$. The difference between $T_{c}^{*}$ from heating and cooling diminishes as $P^*$ increases and vanishes within the errorbars between $P^{*}=0.14$ and $0.145$. Note that above $P^*_c$ no transition exists at all and the estimated $T_{c}^{*}$ is at best a crossover. We thus estimate $P^{*}=0.142(3)$ with $T_{c}^{*}=1.349(2)$.

\begin{figure}[b]
\includegraphics[width=8.5cm]{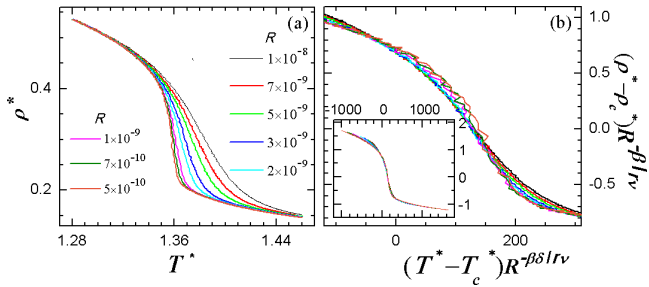}
\caption{\label{density}(Color online) Reduced density versus temperature (a) and its rescaled (b) for various sweep rates listed at $P_{c}^{*}$. Inset: The entire rescaled curves.}
\end{figure}
\begin{figure}
\includegraphics[width=8.5cm]{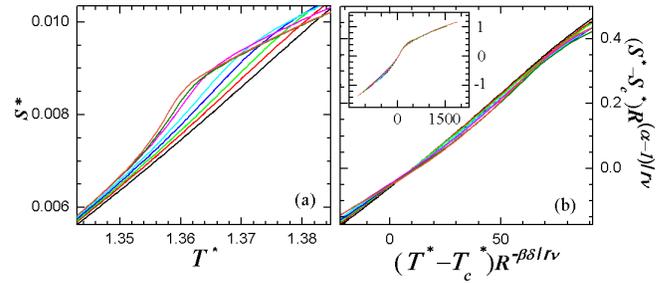}
\caption{\label{entropy}(Color online) Reduced entropy density versus temperature (a) and its rescaled (b) for various sweep rates at $P_{c}^{*}$. Inset: The entire rescaled curves.}
\end{figure}
\begin{figure}[b]
\includegraphics[width=8.5cm]{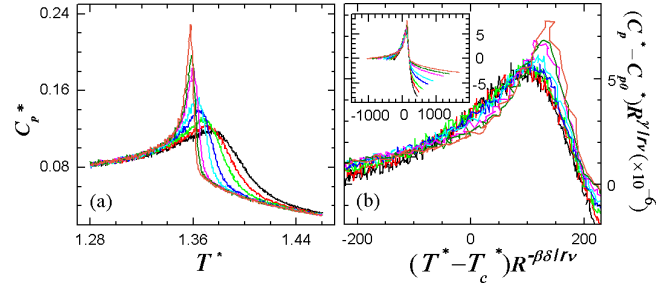}
\caption{\label{cp}(Color online) Reduced isobaric specific heat versus temperature (a) and its rescaled (b) for various sweep rates at $P_{c}^{*}$. Inset: The entire rescaled curves.}
\end{figure}
Having determined $P_c^*$, we then apply the method of iterations to find other parameters. The results are
$\rho_{c}^{*}=0.308(1)$, $\beta/r\nu=0.094(1)$, $T_{c}^{*}=1.348(1)$, and $\beta\delta/r\nu=0.448(5)$ from $\rho^*$ and $S_{c}^{*}=0.0076(1)$, $(1-\alpha)/r\nu=0.253(1)$, $T_{c}^{*}=1.351(2)$, and $\beta\delta/r\nu=0.445(2)$ from $S^*$. Consistencies in $T_{c}^{*}$ and $\beta\delta/r\nu$ are manifest. These results yield reasonably data collapses according to Eqs.~(\ref{drho}) and (\ref{ssca1}) as seen in Figs.~\ref{density} and~\ref{entropy} for the averaged density and entropy, respectively, noting that only the leading contributions have been considered. Using $\gamma=\beta(\delta-1)$, we can also plot data collapses for $C_p^*$ according to Eq.~(\ref{cpsca1}) for a further check. The background is set at $C_{p0}^{*}=0.083$, the common $C_p^*$ of different $R$ at the lowest $T^*$, as $C_{p0}^{*}$ must change little in the vicinity of $T_c^*$. The rescaled curves overlap quite well as shown in Fig.~\ref{cp} and corroborate our results.

All individual critical exponents can be derived from the three known ratios using the scaling laws $\alpha+\beta+\beta\delta=2$ and $\alpha=2-3\nu$. The results are $\delta=4.74(7)$, $\gamma=1.226(28)$, $\beta=0.328(7)$, $\nu=0.627(5)$, $\alpha=0.119(16)$, all in good agreement with $4.789(2)$, $1.233(22)$, $0.3265(3)$, $0.6301(4)$, and $0.110(1)$, respectively, of the Ising universality class~\cite{pelissetto}, and $z=3.078(40)$, also consistent with theoretical~\cite{ohka,sighh,hfb}, experimental~\cite{burssengers,bergzimmerli}, and the recent simulational~\cite{chen} results. In addition, $r=5.556(52)$ and $r_T=4.673(42)$, quite different. This indicates that absence of the complete scaling must lead to far poorer results if only the leading behavior is considered.

The nonuniversal critical parameters obtained agree with $\rho_c^*=0.304(6)$~\cite{smit}, $T^*_c=1.35$~\cite{smit}, and $P_c^*=0.147(2)$~\cite{caillol} individually albeit differing somehow from $\rho_{c}^{*}=0.316(2)$~\cite{okumura,smit,caillol,potoff,lotfi,johnson,puom}, $P_c^*=0.13(1)$ and $T_c^*=1.32(1)$ in the thermodynamic limit~\cite{caillol,potoff00,potoff,lotfi,johnson,puom}. Several factors are of concern. One is the precision of $P_c$. Within the theory, one can show, however,  that in the FTS regime, small deviations of $P_c$ do not modify the results, while too close to $T_c$ due to too small $R$ will lead to a crossover to a regime entirely determined by the deviations, a regime in which the $R$-dependence is only a correction. This is similar to the case where there is a residue magnetic field in magnetic systems~\cite{mask}. Others include contributions from the subleading terms and the small size~\cite{zhong,Huang}. Further studies are needed to clarify their effects. However, they should not change our main conclusion as the relative exponents are divided by the large rate exponent in FTS and are thus small~\cite{xiong,xiong12}.

Summarizing, we have shown that along the critical isobar, the complete scaling results in a unique leading scaling different from the simple and revised scaling. We have also developed a complete-field FTS theory for fluid criticality and a corresponding method to estimate both the static and the dynamic critical exponents as well as critical parameters using molecular dynamics simulations in an isobaric-isothermal ensemble alone.

We greatly appreciate J. V. Sengers, M. A. Anisimov, A. Angell, and S. K. Das for their information and discussions. This work was supported by the NNSF of PRC (Grant No. 10625420).


\begin{thebibliography}{99}\label{sec:TeXbooks}%
%


\bibitem{Andrews}T. Andrews, Phil. Trans. R. Soc. London {\bf 159}, 575 (1869).
\bibitem{bsa}H. Behnejad, J. V. Sengers, and M. A. Anisimov, \emph{Applied Thermodynamics of Fluids}, Chap. 10, PP: 321-339 (Royal Society of Chemistry, 2010).
\bibitem{lrld}L. Cailletet and E. C. Mathias, C.R. Hebd. Seances Acad. Sci. (Paris) {\bf 102}, 1202 (1886).
\bibitem{rm}J. J. Rehr and N. D. Mermin, Phys. Rev. A {\bf 8}, 472 (1973).

\bibitem{Widom70}B.Widom and J. S. Rowlinson, J. Chem. Phys. {\bf 52}, 1670 (1970).
\bibitem{Mermin} N. D. Mermin, Phys. Rev. Lett. {\bf 26}, 169 (1971); {\bf 29}, 957 (1971).


\bibitem{fish00}M. E. Fisher and G. Orkoulas, Phys. Rev. Lett. {\bf85}, 696 (2000).
\bibitem{kfo03}Y. C. Kim, M. E. Fisher, and G. Orkoulas, Phys. Rev. E {\bf 67}, 061506 (2003).
\bibitem{ofp}G. Orkoulas, M. E. Fisher, and A. Z. Panagiotopoulos, Phys. Rev. E {\bf 63}, 051507 (2001).
\bibitem{kf}Y. C. Kim and M. E. Fisher, Phys. Rev. E {\bf68}, 041506 (2003).

\bibitem{yy}C. N. Yang and C. P. Yang, Phys. Rev. Lett. {\bf 13}, 303 (1964).
\bibitem{aniwang}M. A. Anisimov and J. Wang, Phys. Rev. Lett. {\bf 97}, 025703 (2006);  J. Wang and M. A. Anisimov, Phys. Rev. E {\bf75}, 051107 (2007).
\bibitem{Bertrand}C. E. Bertrand, J. F. Nicoll, and M. A. Anisimov, Phys. Rev. E {\bf 85}, 031131 (2012).

\bibitem{Cerd}C. A. Cerdeiri\~{n}a, M. A. Anisimov, and J. V. Sengers, Chem. Phys. Lett. {\bf 424}, 414 (2006).
\bibitem{Wang08} J. Wang, C. A. Cerdeiri\~{n}a, M. A. Anisimov, and J. V. Sengers, Phys. Rev. E {\bf 77}, 031127 (2008).
\bibitem{Perez15}G. P\'{e}rez-S\'{a}nchez, P. Losada-P\'{e}rez, C. A. Cerdeiri\~{n}a, J. V. Sengers, and M. A. Anisimov, J. Chem. Phys. {\bf 132}, 154502 (2010).
\bibitem{MHuang}M. Huang, Z. Chen, T. Yin, X. An, and W. Shen, J. Chem. Eng. Data {\bf 56}, 2349 (2011).
\bibitem{Vale} V. Vale, B. Rathke, S.Will, and W. Schr\"{o}er, J. Chem. Eng. Data {\bf 56}, 1330 (2011).
\bibitem{Perez21}G. P\'{e}rez-S\'{a}nchez, P. Losada-P\'{e}rez, C. A. Cerdeiri\={n}a, and J. Thoen, J. Chem. Phys. {\bf 132}, 214503 (2010).
\bibitem{Losada}P. Losada-P\'{e}rez, C. S. P. Tripathi, J. Leys, C. A. Cerdeiri\~{n}a, C. Glorieux, and J. Thoen, J. Chem. Phys. {\bf 134}, 044505 (2011).
\bibitem{ofu}G. Orkoulas, M. E. Fisher, and C. \"{U}st\"{u}n, J. Chem. Phys. {\bf113}, 7530 (2000).
\bibitem{Wycz}A. K. Wyczalkowska, M. A. Anisimov, J. V. Sengers, and Y. C. Kim, J. Chem. Phys. {\bf 116}, 4202 (2002).

\bibitem{Widom}J. S. Rowlinson and B. Widom, \emph{Molecular Theory of Capillarity}, ch. 6 (Clarendon, Oxford, 1982).
\bibitem{Gubbins}K. E. Gubbins, Mol. Simul. {\bf 2}, 223 (1989).
\bibitem{panagio}A. Z. Panagiotopoulas, Mol. Phys. \textbf{61}, 813 (1987); J. Phys.: Condens. Matter \textbf{12}, R25 (2000).

\bibitem{mf}D. M\"{o}ller and J. Fischer, Mol. Phys. \textbf{69}, 463 (1990); Erratum, {\bf 70}, 1461 (1990).
\bibitem{okumura}H. Okumura and F. Yonezawa, J. Chem. Phys. {\bf 113}, 9162 (2000).
\bibitem{Szalai}I. Szalai, J. Liszi, D. Boda, Chem. Phys. Lett. {\bf 246}, 214 (1995).
\bibitem{okumura01}H. Okumura and F. Yonezawa, J. Phys. Soc. Jpn. \textbf{70}, 1990 (2001).
\bibitem{widom}B. Widom, J. Chem. Phys. \textbf{39}, 2808 (1963).
\bibitem{wilding}N. B. Wilding, J. Phys.: Condens. Matter \textbf{9}, 585 (1997).
\bibitem{wahu}H. Watanabe, N. Ito, and C. K. Hu, J. Chem. Phys. {\bf 136}, 204102 (2012).
\bibitem{fss}M. E. Fisher and M. N. Barber, Phys. Rev. Lett. \textbf{28} 1516 (1972).
\bibitem{Cardyfss}\emph{Finite Size Scaling}, edited by J. Cardy (Amsterdam: North-Holland) (1988).
\bibitem{privman}\emph{Finite Size Scaling and Numerical Simulation of Statistical Systems}, edited by V. Privman (World Scientific, Singapore, 1990).


\bibitem{Wildingb}N. B. Wilding and K. Binder, Physica A {\bf 231}, 439 (1995).


\bibitem{Wilding92}N. B. Wilding and A. D. Bruce, J. Phys.: Condens. Matter {\bf 4}, 3087 (1992).

\bibitem{Bruce}A. D. Bruce and N. B. Wilding, Phys. Rev. Lett. {\bf 68}, 193 (1992).
\bibitem{ferswe}A. M. Ferrenberg and R. H. Swendsen, Phys. Rev. Lett. 61, 2635 (1988); \emph{ibid.}, 63, 1195 (1989).
\bibitem{kim}Y. C. Kim and M. E. Fisher, J. Phys. Chem. B \textbf{108}, 6750 {2004}.


\bibitem{Binderc}K. Binder, Z. Phys. B: Condens. Matter {\bf 43}, 119 (1981).

\bibitem{kfl}Y. C. Kim, M. E. Fisher, and E. Luijten, Phys. Rev. Lett. {\bf 91}, 065701 (2003).
\bibitem{kflc}Y. C. Kim and M. E. Fisher, Comp. Phys. Commun. {\bf169}, 295 (2005).
\bibitem{Kim05}Y. C. Kim, Phys. Rev. E {\bf 71}, 051501 (2005).
\bibitem{hohenberg}P. C. Hohenberg and B. I. Halperin, Rev. Mod. Phys. {\bf 49}, 435 (1977).

\bibitem{ohka}T. Ohta and K. Kawasaki, Prog. Theor. Phys. (Kyoto) {\bf 55}, 1384 (1976).
\bibitem{sighh}E. D. Siggia, B. I. Halperin, and P. C. Hohenberg, Phys. Rev. B {\bf 13}, 2110 (1976).
\bibitem{hfb}H. Hao, R. A. Ferrell, and J. K. Bhattacharjee, Phys. Rev. E {\bf 71}, 021201 (2005).
\bibitem{burssengers}H. C. Burstyn and J. V. Sengers, Phys. Rev. Lett. {\bf 45}, 259 (1980).
\bibitem{bergzimmerli}R. F. Berg, M. R. Moldover, and G. A. Zimmerli, Phy. Rev. Lett. {\bf 82}, 920 (1999); Phys. Rev. E {\bf 60}, 4079 (1999).
\bibitem{chen}A. Chen, E. H. Chimowitz, S. De, and Y. Shapir, Phys. Rev. Lett. {\bf 95}, 255701 (2005).
\bibitem{jy}K. Jagannathan and A. Yethiraj, Phys. Rev. Lett. {\bf 93}, 015701 (2004).
\bibitem{Smith}S.W. Smith, C. K. Hall, and B. D. Freeman, J. Chem. Phys. {\bf 102}, 1057 (1995); J. Comput. Phys. {\bf 134}, 16 (1997).
\bibitem{sm}J. V. Sengers and M. R. Moldover, Phys. Rev. Lett. {\bf 94}, 069601 (2005).
\bibitem{das}S. K. Das, M. E. Fisher, J. V. Sengers, J. Horbach, and K. Binder, Phys. Rev. Lett. {\bf 97}, 025702 (2006); S. K. Das, J. Horbach, K. Binder, M. E. Fisher, J. V. Sengers, J. Chem. Phys., {\bf 125}, 024506 (2006).
\bibitem{deM}E. de Miguel, E. Martin del Rio, and M. M. Telo da Gama, J. Chem. Phys. {\bf 103}, 6188 (1995).
\bibitem{gong}S. Gong, F. Zhong, X. Huang, and S. Fan, New J. Phys. {\bf 12}, 043036 (2010).
\bibitem{zhong} F. Zhong, in \emph{Applications of Monte Carlo Method in Science and Engineering}, edited by S. Mordechai (Intech, Rijeka, Croatia, 2011), p. 469. Available at  http://www.dwz.cn/B9Pe2

\bibitem{huang}X. Huang, S. Gong, F. Zhong, and S. Fan, Phys. Rev. E {\bf 81}, 041139 (2010).
\bibitem{xiong}W. Xiong, F. Zhong, W. Yuan, and S. Fan, Phys. Rev. E {\bf 81}, 051132 (2010).
\bibitem{xiong12}W. Xiong, F. Zhong, and S. Fan, Comp. Phys. Commun. {\bf183}, 1162 (2012).
\bibitem{Huang}Y. Huang, S. Yin, B. Feng, and F. Zhong, Phys. Rev. B {\bf 90}, 134108 (2014).

\bibitem{Yin13} S. Yin, X. Qin, C. Lee, and F. Zhong, arXiv: 1207.1602.
\bibitem{Yin14} S. Yin, P. Mai, and F. Zhong, Phys. Rev. B {\bf 89}, 094108 (2014).
\bibitem{Hu}Q. Hu, S. Yin, and F. Zhong, Phys. Rev. B {\bf 91}, 184109 (2015).

\bibitem{Cardy}J. Cardy, {\it Scaling and Renormalization in Statistical Physics},  (Cambridge University Press, Cambridge, 1996).
\bibitem{zhong06}F. Zhong, Phys. Rev. E {\bf 73}, 047102 (2006).

\bibitem{Chu}B. Chu, F. J. Schoenes, and M. E. Fisher, Phys, Rev. {\bf 185}, 219 (1969).
\bibitem{zhong05}F. Zhong and Q. Z. Chen, \prl {\bf 95,} 175701 (2005); F. Zhong, e-print arXiv1205.1400 (2012).

\bibitem{berendsen}H. J. C. Berendsen \emph{et al.}, J. Chem. Phys. {\bf 81}, 3684 (1984).

\bibitem{frensmit}D. Frenkel and B. Smit, \emph{Understanding Molecular Simulation From Algorithms to Applications}, (Academic Press, Harcourt, 2002).



\bibitem{pelissetto}A. Pelissetto and E. Vicari, Phys. Rep. {\bf 368}, 549 (2002).
\bibitem{smit}B. Smit, J. Chem. Phys. {\bf 96}, 8639 (1992).
\bibitem{caillol}J. M. Caillol, J. Chem. Phys. {\bf 109}, 4885 (1998).
\bibitem{potoff}J. J. Potoff, and A. Z. Panagiotopoulos, J. Chem. Phys. {\bf 109}, 10914 (1998).

\bibitem{lotfi}A. Lotfi, J. Bravec, and J. Fischer, Mol. Phys. \textbf{76}, 1319 (1992).
\bibitem{johnson}J. K. Johnson, J. A. Zollweg, and K. E. Gubbins, Mol. Phys. \textbf{78}, 591 (1993); W. Shi, and J. K. Johnson, Fluid Phase Equilib., \textbf{187}, 171 (2001).
\bibitem{puom}J. Perez-Pellitero, P. Ungerer, G. Orkoulas, and A. D. Mackie, J. Chem. Phys. {\bf 125}, 054515 (2006).
\bibitem{potoff00}J. J. Potoff and A. Z. Panagiotopoulos, J. Chem. Phys. {\bf 112}, 6411 (2000).
\bibitem{mask}N. Goldenfeld, \emph{Lectures on Phase Transitions and the Renormalization Group} (Addison-Wesley, 1992).

\end{thebibliography}
\end{document}